\documentclass[aps,prd,twocolumn,nomerge,superscriptaddress,preprintnumbers,superscriptaddress,nofootinbib,floatfix]{revtex4-1}
\usepackage{stix}
\usepackage{dsfont}
\usepackage{longtable} 
\usepackage{graphicx}
\usepackage{amsmath}
\usepackage{amssymb}
\usepackage{mdwlist}
\usepackage[caption=false]{subfig}
\usepackage{tikz}
\usepackage{siunitx}
\usepackage{placeins}
\usepackage{color}
\usepackage{standalone}
\usepackage{dcolumn}
\usepackage{microtype}
\usepackage{braket}

\usepackage{enumitem}

\usetikzlibrary{patterns}

\newcommand{\beq}{\begin{equation}}
\newcommand{\eeq}{\end{equation}}

\newcommand{\dcc}{LIGO-P1800185}

\graphicspath{{./}}

\begin{document}

\preprint{\dcc}

\title[CW 8 Phase]{Phase decomposition of the template metric for continuous gravitational-wave searches}



\author{S. Mastrogiovanni}
\email{simone.mastrogiovanni@roma1.infn.it}
\affiliation{INFN, Sezione di Roma, I-00185 Roma, Italy.}
\affiliation{Dipartamento di Fisica, Universit\`{a} di Roma ``Sapienza'', P.\
le A.\ Moro, 2, I-00185 Rome, Italy.}

\author{P. Astone}
\affiliation{INFN, Sezione di Roma, I-00185 Roma, Italy.}

\author{S. D Antonio}
\affiliation{INFN, Sezione di Roma Tor Vergata, I-00185 Roma, Italy.}

\author{S. Frasca}
\affiliation{INFN, Sezione di Roma, I-00185 Roma, Italy.}
\affiliation{Dipartamento di Fisica, Universit\`{a} di Roma ``Sapienza'', P.\
le A.\ Moro, 2, I-00185 Rome, Italy.}

\author{G. Intini}
\affiliation{INFN, Sezione di Roma, I-00185 Roma, Italy.}
\affiliation{Dipartamento di Fisica, Universit\`{a} di Roma ``Sapienza'', P.\
le A.\ Moro, 2, I-00185 Rome, Italy.}

\author{I. La Rosa}
\affiliation{INFN, Sezione di Roma, I-00185 Roma, Italy.}

\author{P. Leaci}
\affiliation{INFN, Sezione di Roma, I-00185 Roma, Italy.}
\affiliation{Dipartamento di Fisica, Universit\`{a} di Roma ``Sapienza'', P.\
le A.\ Moro, 2, I-00185 Rome, Italy.}

\author{A. Miller}
\affiliation{INFN, Sezione di Roma, I-00185 Roma, Italy.}
\affiliation{Dipartamento di Fisica, Universit\`{a} di Roma ``Sapienza'', P.\
le A.\ Moro, 2, I-00185 Rome, Italy.}

\author{F. Muciaccia}
\affiliation{INFN, Sezione di Roma, I-00185 Roma, Italy.}
\affiliation{Dipartamento di Fisica, Universit\`{a} di Roma ``Sapienza'', P.\
le A.\ Moro, 2, I-00185 Rome, Italy.}

\author{C. Palomba}
\affiliation{INFN, Sezione di Roma, I-00185 Roma, Italy.}

\author{O.J. Piccinni}
\affiliation{INFN, Sezione di Roma, I-00185 Roma, Italy.}

\author{A. Singhal}
\affiliation{INFN, Sezione di Roma, I-00185 Roma, Italy.}


\date{\today}

\begin{abstract}
A type of gravitational-wave signals in the LIGO-Virgo sensitivity band are expected to be emitted by spinning asymmetric neutron stars, with rotational frequencies that could plausibly emit continuous gravitational radiation in the most sensitive band of the LIGO-Virgo detectors. The most important feature of such kind of signals is in their phase evolution, which is stable over a long observation run. When using analysis based on matched filtering,  the phase evolution of long-coherent signals is needed to define how to build a proper template grid in order to gain the best signal-to-noise ratio possible. This information is encoded in a matrix called \textit{phase metric}, which characterizes the geometry for the likelihood  given by the matched filtering. Most of the times,  the metric for long-coherent signals cannot be computed anlaytically and even its numerical computation is not possible due to numerical precision. In this paper we show a general phase decomposition technique able to make the template metric analytically computable. We will also show how this variables can be employed to distinguish in a robust way among astrophysical  signals and non-stationary noise artifacts that may affect analysis pipelines.
\end{abstract}

\maketitle

%
%
%
%
%

\section{Introduction}
During the first and second advanced LIGO-Virgo \cite{Aasi20152, Acernese2015}  observing runs  several  gravitational-wave signals (GWs) have been detected. The  signals detected so far are short-lived, also called {\it transient}, because their duration is much smaller than the usual observing time of the detectors. In particular five detection from binary black hole merger \cite{Abbott20162,2016PhRvL.116x1103A, 2017PhRvL.119n1101A,2017PhRvL.118v1101A,TheLIGOScientificCollaboration2017} and a detection from a binary neutron star merger \cite{2017PhRvL.119p1101A} were made. The last detection have carried out many astrophysical information on the physics and astrophysics related to neutron stars such as the equation of state \cite{2017PhRvL.119p1101A}. Spinning neutron stars (NSs) are also expected to emit GWs if an asymmetry is present with respect to the rotation axis. These type of signals are expected to be continuous and long-lived with respect to the usual observation time of GW detectors but on the other hand their expected amplitude is very weak. The GW amplitude can carry out useful information on the star's ellipticity and hence on its equation of state.  These type of signals are refereed to as Continuous Waves (CWs).
Generally speaking, algorithms for the detection of CWs match a set of {\it waveform templates} to detectors data, in order to highlight the presence of an astrophysical  signal. For instance, in {\it targeted searches} a wave template covering the entire data set is used, while in {\it semi-coherent searches} a large set of templates is applied  covering smaller portions of the data and later combined incoherently. A  common problem is then how to build the waveform template and to decide what is the spacing into the template lattice. In principle, one should build a grid in the search parameter space such a way that the discretization does not prevent the detection of a signal, and that the computational cost of the entire search is affordable, while exploring a reasonable physical parameter space.
The problem of template spacing has been  originally deeply studied for compact binaries coalescence \cite{PhysRevD.49.2658,PhysRevD.46.5236,PhysRevD.53.6749} and later inspected for CW \cite{PhysRevD.58.063001,Cutler2005,Wette2013,Wette2015,DavidPhD,Brady1997}.
 It has been shown that the information on how to build the template encoded in the so-called template metric \cite{PhysRevD.53.6749,Cutler2005}. As we will see later in Sec. \ref{sec:2}, the metric is represented by matrix defined to compute the signal-to-noise ratio loss given a mismatched template with respect to the one present in the data. This matrix can be used to compute the fraction of the signal-to-noise ratio loss when a discretization error in the template grid is present. On the other hand, the metric carries also information on the of the likelihood function with respect to the  waveform templates. An accurate evaluation of the template metric is then needed in such a way to probe the presence of a CW signal. Unfortunately, for a CW signal the metric is often an ill-conditioned and non-diagonal matrix. The condition number is defined as the ratio of the highest eigenvalue over the smallest eigenvalue of a given matrix. Usually if it is larger than the numerical precision of a compiler ($~10^{16}$) the matrix cannot be inverted properly by algorithms, which makes the template matrix difficult to handle from a numerical point of view. In this paper we will show  a variable decomposition which makes the metric analytically calculable. We will show how to efficiently build a template grid using the new variables and how can we use it to improve CW searches by distinguish a signal from non-gaussian noise by the geometry of the likelihood function. The paper is organized as follows: In Sec. \ref{sec:2} a background for the data analysis will be provided, in Sec. \ref{sec:3} the new variable redefinition will be introduced, in Sec. \ref{sec:4}  tests aimed to probe that the phase decomposition works properly will be shown. Finally in Sec. \ref{sec:5} and Sec.  \ref{sec:6} possible applications for hypothesis testing will be presented, focusing also at the end on implementation in follow-up algorithms.
\section{Data analysis background \label{sec:2}}
In this section we  introduce the data analysis background. We use the  $\mathcal{F}$-statistic defined in \cite{Jaranowski1998}.  The choice of using the $\mathcal{F}$-statistic is due to the fact that we wish to have an estimator that is directly related to the likelihood function. However, our approach is quite general and holds for all searches that use matched filtering technique.

\subsection{The signal model}
The GW signal emitted by an asymmetric  spinning neutron star can be written following the formalism first introduced in \cite{2010CQGra..27s4016A} at the detector reference frame as the real part of
\begin{equation}
h(t)= h_0 f(\eta) \big[ H^+ A_+ (t) + H^\times A_\times (t) \big]e^{2 \pi i f_{\mathrm{gw}} (t) t+i \phi_0},
\label{eq:Hgrande}
\end{equation}
where $f_\mathrm{gw} (t)$ is the GW frequency at the detector reference frame and $\phi_0$ is the phase at the reference time while $f(\eta)$ is a function of the parameter $\eta$. 
For the quadrupole GW emission of rotating tri-axial rigid body (the NS) we expect  $f_\mathrm{gw} (t)$ to be two times the rotational frequency of the spinning neutron star. The polarization amplitudes $H^+, H^\times$ are given by:
\begin{align}
H^+ =\frac{\cos(2 \psi) - i \eta \sin (2 \psi)}{\sqrt{1+\eta^2}},  \, \,  H^\times =\frac{\sin(2 \psi) + i \eta \cos (2 \psi)}{\sqrt{1+\eta^2}},  
\label{eq:HpHc}
\end{align}
with $\eta$ being the ratio of the polarization ellipse semi-minor to semi-major axis and $\psi$ the polarization angle\footnote{It is defined as the direction of the major axis with respect to the celestial parallel of the source measured counter-clockwise.}\cite{2010CQGra..27s4016A}. The detector \textit{sidereal responses} to the GW polarizations are encoded in the functions $A_+ (t), A_\times (t)$. It can be shown that the waveform defined by Eq. \eqref{eq:Hgrande} is equivalent to the GW signal expressed in the more standard formalism of \cite{Jaranowski1998}, given the following relations: 
\begin{equation}
\eta=-\frac{2\cos \iota}{1+\cos^2 \iota},
\label{eq:etaiota}
\end{equation}
where $\iota$ is the angle between the line of sight and the star rotation axis, and
\begin{equation}
H_0=h_0 \sqrt{ \frac{1+6 \cos ^2 \iota + \cos^4 \iota}{4}},
\end{equation}
with the usual GW amplitude
\begin{equation}
h_0=\frac{1}{d} \frac{4 \pi^2 G }{c^4} I_\mathrm{zz} f_\mathrm{gw}^2 \epsilon,
\label{eq:GW_amplitude}
\end{equation}
where $d,I_\mathrm{zz}$ and $\epsilon$ are respectively the star distance, its moment of inertia with respect to the rotation axis and the {\it ellipticity}, which measures the star degree of asymmetry.  In the detector reference frame, the signal  is  not monochromatic, i.e. the frequency $f_\mathrm{gw} (t)$ in Eq. \eqref{eq:Hgrande} is a function of the different modulation that act on the signal. In fact, the signal is  modulated by several effects, namely the \textit{R\"{o}mer delay} due to the detector motion in the Solar System Barycenter and the source intrinsic spin-down due to the rotational energy loss. The phase modulation of a CW signal can be expressed as a composition of the listed effects \footnote{We are neglecting other phase modulations such as the Einstein, the Shapiro delay \cite{2017CQGra..34m5007M} and the contribution from further derivatives in the NS's frequency Taylor expansion. These modulations are taken into account during the analysis but their effect is neglected when computing the metric since they have a negligible effect on the mismatch (defined later) with respect to the other effects.}:
\begin{align}
\frac{\phi_{\rm gw}(t)-\phi_0}{2 \pi}= &f_{ 0} (t-t_0) + \frac{1}{2}\dot{f}_{0}(t-t_0)^2+ \nonumber \\ & +f_{ 0} \, \vec{\mathcal{P}}(t) \cdot \widehat{n} + \dot{f}_{0}(t-t_0)\,  \vec{\mathcal{P}}(t)\cdot \widehat{n}
\label{eq:phevol}
\end{align}
where $t_0$ is a reference time, $\vec{\mathcal{P}}(t)$ the position of the Earth in the Solar System Barycenter (normalized to the speed of light), $\phi_0$ an initial phase and $\widehat{n}$ the versor pointing to the source location in the sky. The variables that determine the phase evolution in Eq. \eqref{eq:phevol} are the GW frequency $f_0$ and it's derivative $\dot{f_0}$ (at a given reference time)  together with two angular variables which are the position in the sky $\alpha,\delta$. For the sky-position we will use the equatorial coordinates.  With the signal description presented in Eq. \eqref{eq:Hgrande} the signal naturally factorizes as the product of some GW amplitudes $H_p(\vec{\beta_s})$ which are complex scalar numbers, and depend on the so-called \textit{extrinsic parameters} $\vec{\beta_s}=(\eta, \psi, \phi_0)$, and phase templates $\ket{\mathcal{A}_p(\vec{\lambda_s})}$ which are vectors and depend on the the so-called {\it intrinsic parameters} (or phase parameters) $\vec{\lambda}_s=(\alpha,\delta, f, \dot{f})$. 
Without loosing of generality we can use the braket notation to indicate that the templates (or data itself) can either be expressed in  different basis such as the frequency basis (Fourier's domain) or time basis (time series). A signal composed by $p$ polarizations  can be generally written as \cite{Jaranowski1998,2010CQGra..27s4016A}
\begin{equation}
\ket{h}=H_0 (h_0,\eta) \sum_p H_p(\vec{\beta}_s) \ket{\mathcal{A}_p(\vec{\lambda)}_s}.
\label{eq:bra}
\end{equation}
For instance, the phase templates $\ket{\mathcal{A}_p(\vec{\lambda)}_s}$ can be  expressed in the time domain (using the time basis $\widehat{t}$) as harmonic functions. In fact they will be the product of the detector sidereal responses $A_{+/\times} (t, \vec{\lambda}_s)$ and all the possible phase modulations of the signal.
\begin{equation}
\braket{\widehat{t}|\mathcal{A}_p}=A_{p} (t ,\vec{\lambda}_s) e^{i \phi_{\rm gw}(t, \vec{\lambda}_s)}
\label{eq:tb}
\end{equation}

\subsection{Definition of the statistic}
Following the same approach as in \cite{Jaranowski1998} we can model our data as the superposition of  Gaussian noise and a possible signal
\begin{equation}
\ket{x}=\ket{n}+\ket{h}
\label{eq:super}
\end{equation} 
The likelihood for the data $x$  containing a signal  $h$ can be expressed as:
\begin{equation}
\mathcal{L}(x|h(\vec{\lambda})) \propto e^{\frac{1}{2} \braket{x-h|x-h}}.
\end{equation}
The inner scalar product can be performed both in time domain or in the frequency domain:
\begin{equation}
\braket{a|b}= \frac{2}{S_{f}} \int_0^{f_{max}} a(f) \cdot b^*(f) df =\frac{2}{S_{f}} \int_0^{T_{\rm coh}} a(t) \cdot b^*(t) dt
\end{equation}
with  $S_f$ being the unilateral detector noise spectrum that since we are looking at a very narrow-frequency region (order of $10^{-3}$ Hz) we assume to be constant , $T_{\rm coh}$ the coherent integration time of the analysis and ``$^*$'' the complex conjugation.  One can assume $S_f$ to be almost constant over a small frequency band in the  case of nearly gaussian noise.  We then define the maximum likelihood estimator as the ratio of the likelihoods associated to a signal being present and not, respectively: 
\begin{equation}
\mathcal{L}_{\rm ML}=\frac{\mathcal{L}(x|h)}{\mathcal{L}(x|h=0) }
\label{eq:ML}
\end{equation}
The signal detection problem consists to maximize  Eq. \eqref{eq:ML} while trying many different signal templates  which are functions as well  of the extrinsic and intrinsic GW parameters, $\vec{\lambda}$ and $\vec{\beta}$.
It can be shown that $\mathcal{L}_{\rm ML}$ can be analytically maximize with respect to the extrinsic parameters of the wave thus removing 4 dimensions from our maximization problem. Following the same procedure of \cite{Jaranowski1998} implemented for the signal description in \cite{2010CQGra..27s4016A} it is possible to show that:
\begin{equation}
\mathcal{L}_{\rm ML} = e^{\overset{_*}{\mathcal{F}}},
\label{eq:RML}
\end{equation}
where $\overset{_*}{\mathcal{F}}$ is the $\mathcal{F}$-statistic computed from complex data instead of real data.
\begin{equation}
\overset{_*}{\mathcal{F}}=\frac{1}{2} \sum_p \frac{\braket{\mathcal{A}_p|x}}{\braket{\mathcal{A}_p|\mathcal{A}_p}}\braket{x|\mathcal{A}_p}.
\label{eq:R}
\end{equation}
Theoretically $\overset{_*}{\mathcal{F}}$ has the same statistical meaning of the usual $\mathcal{F}$-statistic, i.e. is the logarithm of the maximum likelihood estimator in Eq. \eqref{eq:ML}. Practically, in order to compute $\overset{_*}{\mathcal{F}}$ the application of two templates is required, while for the usual $\mathcal{F}$-statistic the application of 4 templates is required. This is because $\overset{_*}{\mathcal{F}}$ is computed starting from a complex representation of the data.

The extrinsic parameters can be estimated from the complex polarization amplitude estimators,which result from the application of two matched filters: 
\begin{align}
& \widehat{H}_+=\frac{\braket{A_+|x}}{\braket{A_+|A_+}} \,
& \widehat{H}_\times=\frac{\braket{A_\times|x}}{\braket{A_\times|A_\times}}. 
\label{eq:estimators}
\end{align}
Using the relations given by  Eq. \eqref{eq:HpHc}, one can use the estimators $\widehat{H}_{+/\times}$ of the GW polarization amplitudes to recover the intrisc parameters of the wave a posterior as  a combination of the two estimators in Eq. \eqref{eq:estimators}. The relations can be found in \cite{2017CQGra..34m5007M}.

\subsection{Phase metric}
Let us assume that we are computing the $\overset{_*}{\mathcal{F}}$ statistic by using phase templates calculated using parameters with a small  mismatch $\Delta \vec{\lambda}=|\vec{\lambda}-\vec{\lambda_s}|$ with respect to the signal parameters values. We expect that, depending on $\vec{\Delta \lambda}$, only a fraction of the signal  will be recovered, ideally the one corresponding to the signal if $|\vec{\Delta \lambda}|=0$. The function that quantifies the loss for a mismatch in the phase parameters is called {\it mismatch function}:
\begin{equation}
m_f=\frac{E[\overset{_*}{\mathcal{F}}_{\rm s}]-E[\overset{_*}{\mathcal{F}}_{\rm m}( \vec{\lambda_s}+\Delta \vec{\lambda})]}{E[\overset{_*}{\mathcal{F}}_{\rm s}]}
\label{eq:mism1}
\end{equation}
where $E[\overset{_*}{\mathcal{F}}_{\rm s}]$  and $E[\overset{_*}{\mathcal{F}}_{\rm m}(\Delta \vec{\lambda})]$ are respectively the expected values of the $\overset{_*}{\mathcal{F}}$ statistic for a perfect matched template and for a template computed with a  mismatch $\Delta \vec{\lambda}$.
One can Taylor expand the term $E[\overset{_*}{\mathcal{F}}_{\rm m}]$ around the signal true parameters obtaining the form \cite{PhysRevD.53.6749}: \footnote{The first derivative term is null since we are expanding around a local maximum.}

\begin{eqnarray}
m_f=\sum_{i,j} g_{ij} (\vec{\lambda}_s) \Delta \lambda^i \Delta \lambda^j + \mathcal{O}(\Delta |\vec{\lambda}|^3)
\label{eq:mism}
\end{eqnarray}
The $4\times4$ tensor $g_{ij}(\vec{\lambda}_s)$ represent the  metric in the 4-dimensional parameter space, and $(\vec{\lambda}_s)$ are the signal phase parameters. It is possible to show that if one assumes that the phase displacement due to the sidereal Earth motion is smaller than the phase displacement due to other effects (such as Doppler modulation), the metric will assume the form\cite{PhysRevD.53.6749,Brady1997,DavidPhD}, more details on how to recover Eq. \eqref{eq:phase_metric} using the formalism introduced in Sec. \ref{sec:2}, are also given in Appendix \ref{ap:A}:
\begin{align}
g_{ij}  = & \frac{1}{T_{\rm coh}} \int_0^{T_{\rm coh}}  \frac{\partial \phi_{\rm gw}}{\partial \lambda_i} \frac{\partial \phi_{\rm gw}}{\partial \lambda_i}  \bigg|_{\lambda=\lambda_s} dt+ \ldots 	\nonumber \\ &- \frac{1}{T^2_{\rm coh}}  \int_0^{T_{\rm coh}}   \frac{\partial \phi_{\rm gw}}{\partial \lambda_i}  \bigg|_{\lambda=\lambda_s} dt  \int_0^{T_{\rm coh}}   \frac{\partial \phi_{\rm gw}}{\partial \lambda_j}  \bigg|_{\lambda=\lambda_s} dt.
\label{eq:phase_metric}
\end{align}
The concept of metric can also be extended to semi-coherent searches (see Appendix \ref{ap:B} for more details) or in the case of pulsars in binary systems as done by \cite{PhysRevD.91.102003}. The metric indicates the fraction of signal-to-noise ratio that we are able to recover while searching from a mismatched template. Solving Eq. \eqref{eq:mism} for a constant mismatch $m_f$ with respect to the variables $\Delta \lambda_i$  is equivalent for determine the set of templates which will result in the same value of the $\mathcal{F}$-statistic. Hence it is equivalent of study the hyper-surfaces at constant likelihood with respect to the templates.
So a study of the  metric is very important to uderstand how to build a proper template grid and which are the shape of the likelihood function for a GW signal present ideally only in gaussian noise.  However, as it is possible to understand looking at the phase evolution in Eq. \eqref{eq:phevol} and to the metric in Eq. \eqref{eq:phase_metric}, the computation of the matrix is  not an easy task. The first problem is that the metric is not flat, i.e. every component depends on the signal true parameters, that we do not know. Mathematically this effect arises from the fact that different phase modulations in Eq. \eqref{eq:phevol}  couple one to each other (e.g. the frequency phase evolution and the Doppler modulation). Another problem is that usually the metric is ill-conditioned, i.e. has a condition number higher than the dobule float numerical precision and the computation of the eigendirections (which gives information on the geometry of the likelihood function) can present numerical problems \cite{Wette2013,Prix2007}.
\begin{figure*}[htp!]
\centering
\includegraphics[width=1.0\textwidth]{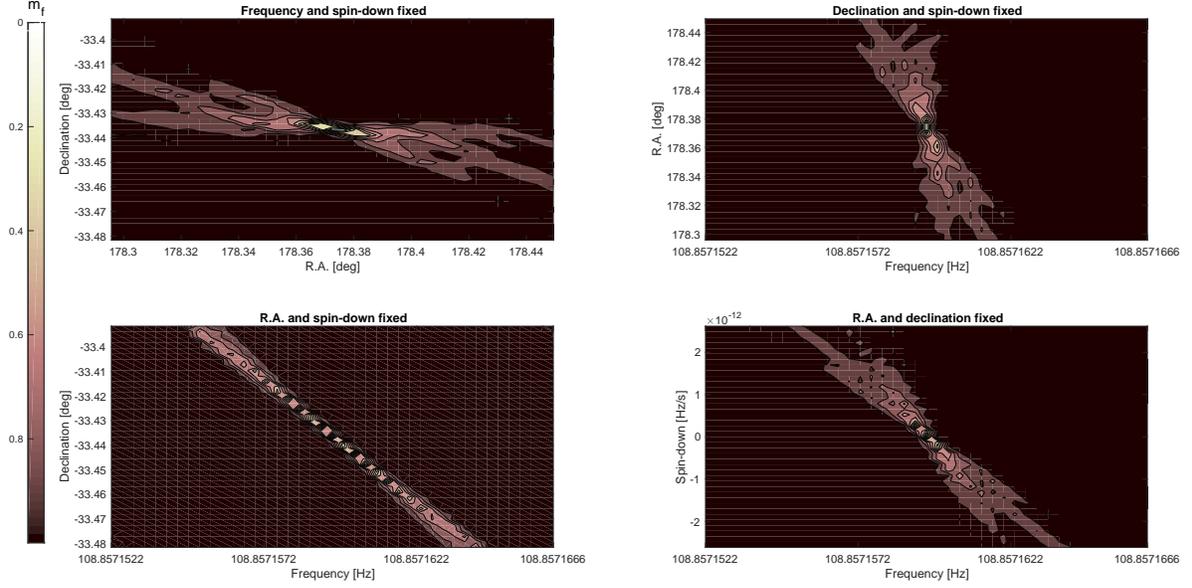}
\caption{Contour plots of the mismatch function $m_f$ looking for an hardware injection with SN R$~70$ in 1 month of O1 data. In every plot, two of the four phase variables ($f, \dot{f},\alpha,\delta$) are fixed at their injected value thus reducing the dimensionality of the problem, while the other two are left free. Reference time $t_0$ in all the plots set at the beginning of the run.}
\label{fig:mmovar}
\end{figure*}
  
As an example on how to compute the metric spacing let us consider a simple case. In narrow-band or directed  searches one assumes the sky-position to be perfectly known, while the frequency $f_0$ and  spin down$\dot{f}_0$  are known with some uncertainties. This is the same case as the one considered in a direct search aimed to look for CW from the centrlal compact object in Cassiopea A\cite{0264-9381-25-23-235011}, where the tempalte spacing was decided using the template metric \cite{PhysRevD.53.6749} for the CW case \cite{DavidPhD}.
 If one assume to correct the phase modulations related to the sky-position in a way that does not depend on $f_0,\dot{f}_0$, like with the non-uniform resampling technique \cite{2017CQGra..34m5007M}, the remaining phase evolution of the signal on the corrected data will be:
\begin{equation}
\frac{\phi(t)-\phi_0}{2 \pi}=f_0(t-t_0)+\frac{1}{2}\dot{f}_0(t-t_0)^2
\end{equation} 
At this level, the metric will be a $2 \times 2 $ tensor, if we compute the metric using Eq. \eqref{eq:phase_metric} and using $f_0$ and $\dot{f}_0$ as variable once can check  that by placing the reference time in the middle of the run, $t_0=T_{\rm coh}/2$:
\begin{equation}
g_{ij}=
  \begin{bmatrix}
    g_{ff} & g_{f\dot{f}}  \\
    g_{\dot{f}f} & g_{\dot{f} \dot{f}} 
  \end{bmatrix}
  \approx
  \begin{bmatrix}
    T_{\rm coh}^2 & 0  \\
    0 & T_{\rm coh}^4 
  \end{bmatrix}
\end{equation}

The metric obtained in this way is already diagonal and the mismatch function can be written as:
\begin{equation}
m_f \approx T^2_{\rm coh} \Delta f^2+ T^4_{\rm coh} \Delta \dot{f}^2
\end{equation}
It is then natural to define the frequency and spin-down resolution as $\Delta f=1/T_{\rm coh}$ and $\Delta \dot{f}=1/T_{\rm coh}^2$ respectively. These are the usual frequency and spin-down  {\it ``bins'}' used in target and narrow-band searches. From the components of the narrow-band metric it is also possible to see that the condition number scales as  the ratio of the highest eigenvalue and the lower one ( $\propto T_{\rm coh}^2$). In the narrow-band case the computation  of the metric  can be done analytically, and the matrix is already diagonal overcoming the numerical inversion of the metric.
However the general case is not  so trivial, the condition number will scale at least with $ T_{\rm coh}^4$, and the surfaces at constant mismatch will not have a trivial shape.  Fig. \ref{fig:mmovar} shows the contour plots of the mismatch function $m_f$ for the likelihood surface around an hardware injection (which are fake signal injected in the experiment for testing purposes) in the first Advanced LIGO observation run (O1).  The injection had a signal-to-noise ratio (SNR) \footnote{Using the same notation introduced in  \ref{sec:2} , given a signal $\ket{h}$, the SNR can be defined as $${\rm SNR}=\sqrt{\braket{h|h}}$$} of  about 70 for one month of data integrated coherently, it is located at $108.85$ Hz and it had an almost zero spin-down.
The plots in Fig. \ref{fig:mmovar} have been generated computing the mismatch by fixing two of the four CW intrinsic variables to their injected value,  we see that even if we have a perfect knowledge on two of the phase variables the problem of templates placing is not so trivial to understand due to the shape of the likelihood surface.  The templates that lie on the patterns showed in Fig. \ref{fig:mmovar} are called ``non-orthogonal'' since they recover fraction of the same signal. The metric is used to compute the distance between two templates in this parameter space, templates that lies on the same pattern will be close each other with respect to the templates that are outside a given pattern.
The general case, in which all the four phase variables are unknown will be more complicated and the likelihood hyper-surfaces difficult to study. 

For these reasons the authors in \cite{Wette2013} have introduced the so-called {\it Super-sky metric}. The idea of such metric is to linearize the phase evolution of the CW signal by relaxing the constrain on the sky-versor, i.e. its components are left free to span in the volume of a {\it 3-d} sphere. After that  a new set of sky-positions $n_a,n_b,n_c$ and frequency parameters $\nu, \dot{\nu}$ which nearly diagonalize the metric (the mixed components of the rotational parameters are non-null).  However, as we will see later, this choice adds one extra-dimension to the templates thus meaning that there is the possibility to explore non-physical templates. This problem has been solved in \cite{Wette2013} by realizing that, once the sky-position and rotational parts of the metric had been decorrelated from each other, the metric dependence on one of the new sky-positions is much smaller than the other two, meaning that the problem is collapsed again on a 2 dimension surface instead of a volume. Our approach, which will be presented in the next section, share the idea of the super-sky metric of adding extra dimensions. However, while the super-sky metric was aimed to build a physical template grid for GW searches, our approach is aimed to probe the presence of the signal inside the data with respect to general phase modulations that are to data, during the analysis.

\section{Maximum Phase decomposition \label{sec:3}}
As pointed out in the previous section our task is to find a method  to make the metric flat, with a reasonable condition number and possibly analytically computable (no dependence on the signal parameter).  Our approach will try to extend the concept of adding extra dimensions to linearize the phase of a CW signal. In this section we  show the logical steps or our approach.

\subsection{Definition and metric computation \label{sec:sa}}
Theoretically one can obtain a flat and analytically computable metric by linearizing the CW phase with respect to each variable, our first step is to decompose the CW phase evolution in Eq. \eqref{eq:phevol} in the following way:
\begin{align}
\frac{\phi_{\rm gw}(t)- \phi_0}{2 \pi}= &f_{ 0} (t-t_0) + \frac{1}{2}\dot{f}_{0}(t-t_0)^2+ \nonumber \\ & +f_{ 0} \, \mathcal{P}_x(t) \cdot n_x +f_{ 0} \, \mathcal{P}_y(t) \cdot n_y+ \nonumber\\ &+   f_{ 0} \, \mathcal{P}_z(t) \cdot n_z + \dot{f}_{0}(t-t_0)\,  \mathcal{P}_x (t)\cdot n_x + \nonumber \\ +& \dot{f}_{0}(t-t_0)\,  \mathcal{P}_y(t) \cdot n_y + \dot{f}_{0}(t-t_0)\,  \mathcal{P}_z(t) \cdot n_z
\label{eq:phevol1}
\end{align}
where we have exploited the components of each scalar product related to the Doppler modulation. The next intent is to write Eq. \eqref{eq:phevol1}  in the form:
\begin{equation}
\phi (t)= \sum_{i=1}^{8} \varphi_i v_i(\tau)
\label{eq:mpd}
\end{equation}
where $v_i(\tau)$ are functions of an adimensional time $\tau=(t-t_0)/T_{\rm obs}$ ($T_{\rm obs}$ is the observation time of the detector) and the variables $\varphi_i$  are a new set of coordinates defined from the usual CW phase parameters $f,\dot{f},\alpha,\delta$. By looking at Eq. \eqref{eq:mpd} and exploiting the products in Eq. \eqref{eq:phevol1}, one can write the new  scalar variables $\varphi_i$ as:
\begin{subequations}
\label{eq:2}
\begin{align}
& \varphi_1=2 \pi f_{0} T_{\rm obs}   \\
& \varphi_2=\pi  \dot{f}_{0} T^2_{\rm obs}  \\
& \varphi_3= 2 \pi  f_{0}  {\rm max}_t[|\mathcal{P}_x(t)|] \cos \alpha \cos \delta \\
&\varphi_4= 2 \pi  f_{0}   {\rm max}_t[|\mathcal{P}_y(t)|] \sin \alpha \cos \delta \\
& \varphi_5= 2 \pi  f_{0}   {\rm max}_t[|\mathcal{P}_z(t)|] \sin \delta\\
&  \varphi_6= 2 \pi  \dot{f}_{0}  T_{\rm obs} {\rm max}_t[|\mathcal{P}_x(t)|]   \cos \alpha \cos \delta \\
&  \varphi_7= 2 \pi  \dot{f}_{0}  T_{\rm obs} {\rm max}_t[|\mathcal{P}_y(t)|] \sin \alpha \cos \delta \\
& \varphi_8= 2 \pi  \dot{f}_{0}  T_{\rm obs} {\rm max}_t[|\mathcal{P}_z(t)|] \sin \delta \\
\end{align}
\end{subequations}
 and the adimensional functions $v_i(\tau)$ as:
\begin{subequations}
\label{eq:3}
\begin{align}
& v_1=\tau  \nonumber  \\
& v_2= \tau^2  \nonumber  \\
&v_3=\mathcal{P}_x(\tau)/{\rm max}_\tau[|\mathcal{P}_x(\tau)|]  \\
&v_4=\mathcal{P}_y(\tau)/{\rm max}_\tau[|\mathcal{P}_y(\tau)|] \\
& v_5=\mathcal{P}_z(\tau)/{\rm max}_\tau[|\mathcal{P}_z(\tau)|] \\
&v_6= \tau \mathcal{P}_x(\tau)/{\rm max}_\tau[|\mathcal{P}_x(\tau)|] \\
& v_7= \tau \mathcal{P}_y(\tau)/{\rm max}_\tau[|\mathcal{P}_y(\tau)|] \\
& v_8= \tau \mathcal{P}_z(\tau)/{\rm max}_\tau[|\mathcal{P}_z(\tau)|] \
\end{align}
\end{subequations}
The new variables defined in Eqs. \eqref{eq:2} represent the maximum phase displacement that a signal may experience during the observing time $T_{\rm obs}$ from the modulation of different physical effects. We called these new decomposition as {\it maximum phase decomposition}. On the other hand the adimensional time functions $f_i (\tau)$ in Eqs. \eqref{eq:3} represent the time evolution of different phase modulations. For instance, the intrinsic frequency phase evolution and spin-down evolution are represented by $\varphi_1$ and $\varphi_2$. The Doppler coupling with the frequency is represented by $\varphi_{3-5}$, while the Doppler coupling with the spin-down is represented by $\varphi_{6-8}$. The values related to the Doppler modulation have 3 components because the Doppler can be decomposed on the usual cartesian coordinates $x,y,z$. Using this new set of variables the metric in Eq. \eqref{eq:phase_metric} can be analytically computed and assumes a very simple form:
\begin{equation}
g^{\varphi}_{ij}=\int_0^1 v_i(\tau) v_j(\tau) d \tau - \int_0^1 v_i(\tau)  d \tau \int_0^1 v_j(\tau) d \tau 
\label{eq:phm}
\end{equation}
where $(i,j=1,\ldots,8)$. Since the time is now adimensional, the integration in Eq. \eqref{eq:phm} go from `` $0$'' which correspond to the start of the run, to ``$1$''  which correspond to the end of the run.   Another advantage of using the maximum phase decomposition is that  the condition number of the metric is naturally constrained and depends  weakly on the amount of data that we are using. This is because we are normalizing each phase component by the maximum phase displacement that can occur during the analysis and the integral in Eq. \eqref{eq:phase_metric} is constrained. A drawback of using 8 variables instead of 4 is the increasing cost of the analysis due to the fact that we are now handling an eight-dimensional parameters. Moreover, since we are extending the dimensionality of the parameter space, not all the templates in the eight-dimensional parameter space will correspond to a template in the four-dimensional CW space (refer to Appendix \ref{ap:C} for more details). However this is a trade that we can afford if the variables are used for hypothesis testing, as we will see later or with Markov chain Monte Carlo techniques, which optimally scale with the dimensionality of the problem. 
\subsection{Diagonalization of the metric \label{sec:sb}}

Even though the condition number is naturally constrained by the maximum phase decomposition, it can be still high such as $10^{10}$ as can be seen in Fig. \ref{fig:cn2}. Even though such kind of condition number can be handled from a double-precision float precision compiler, we would like to perform some {\it ad-hoc} transformation on the metric that further decrease its value. The first step is to express the phase metric (which is now a 8 x 8 symmetric matrix) in Eq. \eqref{eq:phm} as the  product of two square matrix.

\begin{equation}
g^\varphi_{ij}=\sqrt{g^\varphi_{ij}} \sqrt{g^\varphi_{ij}}.
\label{eq:ormetric}
\end{equation}

This procedure  is performed using using the Block-Schur  algorithm \cite{10.1007/978-3-642-36803-5_12}  that does not involve inversions of any kind. The two square root matrixes will have a condition number that is roughly the square root of the original condition number, say $10^5$. After that we factorize the square root matrix using the QR decomposition\cite{Matrix1}: \footnote{The decomposition is unique if the matrix is symmetric and positive defined, true condition for a metric.} 

\begin{equation}
\sqrt{g^\varphi_{ij}}=Q R.
\end{equation}
Where $Q$ is a orthogonal unitary matrix and $R$ is an upper triangular matrix. Rewriting the original metric in Eq. \eqref{eq:ormetric} and doing some linear algebra one can write.

\begin{equation}
g^\varphi_{ij}=R^T Q^T Q R= R^T \mathds{1} R,
\label{eq:R}
\end{equation}
where we have used the fact that $Q$ is orthonormal. From Eq. \eqref{eq:R}  one can  understand that R is the matrix of coordinate transformation that brings from the physical variables $\varphi$ to some phase variables $\Phi$ in which the metric $g_{ij}^{\Phi}$ is a unitary matrix, i.e. all the eigenvalues of the metric $g^{\Phi}_{ij}$  are  ones.  Using this new set of variables, we can easily compute  the mismatch in Eq. \eqref{eq:mism} as a  summation of  quadratic phase displacement.

\begin{equation}
m_f=\sum_{i=1}^8 \Delta \Phi_i^2
\end{equation}
The new variables are measured in radiants. From the above Eq. we see that a mismatch of $|\vec{\Delta \Phi}|=1$ rad will correspond to a mismatch of $m_f=1$. So it's natural to use the $\Phi$ variables to directly measure the distance between two templates. It is also worth to note that using the $\Phi$ variables, the mismatch, and hence the likelihood function, will have spherical symmetry with respect to the true parameter of the signal.

Summarizing in order to obtain the $\Phi$ basis: 
\begin{enumerate}[label=(\roman*)]
\item We use the CW variables $\vec{\lambda}=(f_0,\dot{f_0},\alpha,\delta)$ to define the maximum phase displacements during the analysis given by Eqs . \eqref{eq:2}-\eqref{eq:3}.
\item We compute the metric $g_{ij}^\varphi$ using Eq. \eqref{eq:phm}.
\item We diagonalize $g_{ij}^\varphi$ using the numerical procedure described in Sec. \ref{sec:sb}.
\end{enumerate}
\section{Testing \label{sec:4}}
 In this section we will present the results of several tests aimed to show that the maximum phase decomposition presented in Sec. \ref{sec:sa}, and the diagonalization process presented in Sec. \ref{sec:sb}, properly work  while addressing the problems related to diagonalization. In the next paragraphs we will show tests aimed to check the condition number (which is important to quantify if we can use numerical algorithms to switch from the $\varphi$ variables to the $\Phi$ variables), and the mismatch length in the template parameter space estimated by the metric $g_{ij}^\Phi$. Finally we will also probe if the mismatch predicted by the new metric metric $g_{ij}^{\Phi}$ in the case of a signal is consistent with Eq. \eqref{eq:mism}.

\subsection{Condition Number} 
\begin{figure}[h!]
\includegraphics[width=0.45\textwidth]{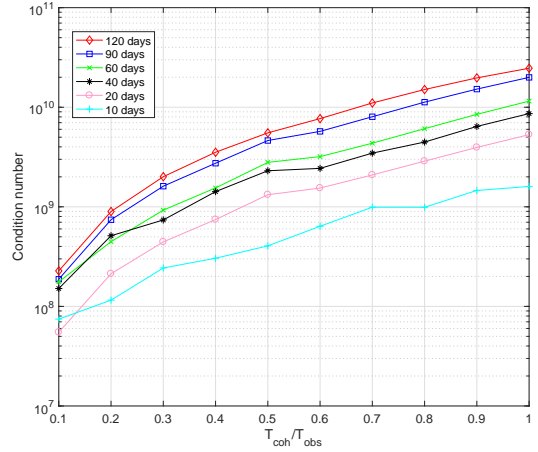}
\caption{Condition number of the metric $g_{ij}^\varphi$ computed with the maximum phase decomposition on the vertical axis. The fraction of data that we are integrating coherently is on the horizontal axis. The lines indicate the condition number computed for different observation times.}
\label{fig:cn}
\end{figure}

\begin{figure}[h!]
\includegraphics[width=0.45\textwidth]{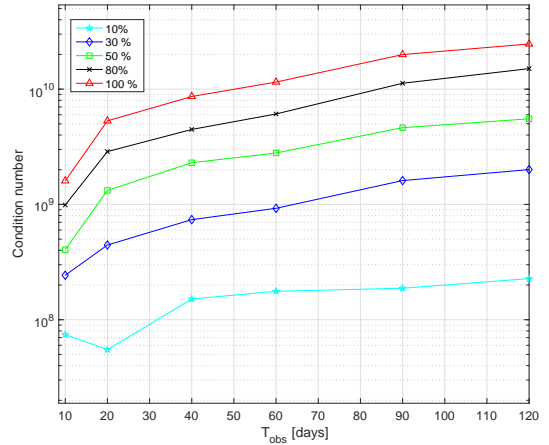}
\caption{Condition number of the metric $g_{ij}^\varphi$ computed with the maximum phase decomposition on the vertical axis with respect to the observation time of the analysis on the horizontal axis. The lines represent the fraction of data we are  coherently analyzing.}
\label{fig:cn2}
\end{figure}

The very first check  is to control the condition number of the metric in the  phases $g_{ij}^\varphi$. As pointed out in \cite{Wette2013} the condition number  increases with the length of data that we are coherently analyzing. This happens because the eigenvalues of the matrix, and then the determinant of the template metric becomes smaller and smaller i.e. a finer template grid is needed. The Maximum phase decomposition is supposed to constrain the condition number to the value corresponding of analyzing full coherently the data set. The plot in Fig. \ref{fig:cn}   reports the value of the condition number as function of the fraction of data that we are integrating coherently. The points with $T_{\rm coh}/T_{\rm obs}=1$ represent a full coherent search, while the points with $T_{\rm coh}/T_{\rm obs}<1$  represent semi-coherent searches.  The condition number is constrained to less than $10^{11}$ which is lower than the double-precision float precision $10^{16}$. Figure \ref{fig:cn} also shows that the condition number increases with the observation time. However, if we plot our results with respect to the observation time,  Fig. \ref {fig:cn2}, we can see that the condition number has a weak scaling with respect to the observation time of the analysis. In the case of a full coherent search (maximum condition number) the value is still constrained below to $10^{11}$. 

From this test we can see that the maximum phase decomposition is properly regularizing the condition number by constraining its value to $10^{11}$ (which can be handled from a double-precision float precision compiler) and making the algorithm able to handle $g_{ij}^\varphi$. Another type of test is to check if the matrix decomposition in Eq. \eqref{eq:R} approximates well the metric in Eq. \eqref{eq:phm}.  We have hence computed the maximum relative error on the estimated phase metric as

\begin{equation}
M={\rm max}_{ij} \big[ \frac{g^\varphi_{ij}-R^T Q^T Q R}{g^\varphi_{ij}}\big].
\label{eq:mm}
\end{equation}

\begin{figure}[h!]
\includegraphics[width=0.45\textwidth]{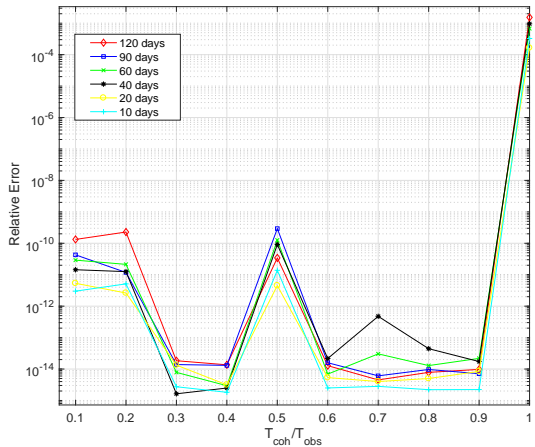}
\caption{Vertical axis: Maximum relative error computed using Eq. \eqref{eq:mm}. Horizontal Axis: Fraction of data used coherently with respect to the observation time. The different types of lines indicate different observation times.}
\label{fig:cn3}
\end{figure}

Figure \ref{fig:cn3} shows the maximum relative error computed as a function of the coherence time we are using in our analysis. Also here the endpoint of the figure represents a full coherent search while the others are semi-coherent search s. It is possible to note that the endpoint of every simulation has a much higher relative error. This is because in the original matrix, $g^\varphi_{23}$\footnote{$g^\varphi_{23}$ correspond to $g_{f \dot{f}}$ of the narrow-band search example in Sec. \ref{sec:2}, hence for a reference time in the middle of the run is  equal to 0.} is almost zero. In conclusion the metric seems to be well-decomposed with the QR decomposition.

\subsection{Mismatch length} 
\begin{figure}[h!]
\includegraphics[width=0.45\textwidth]{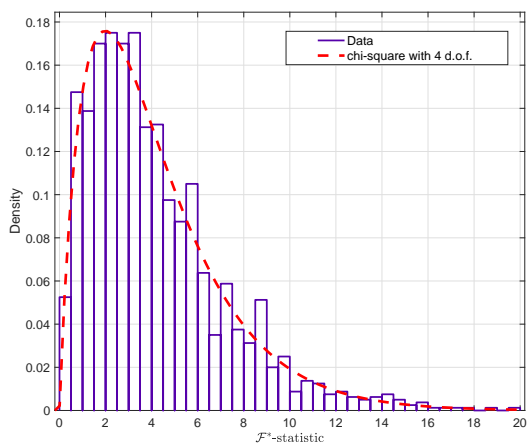}
\caption{Histogram of the detection statistic obtained for a full-coherent search using a 8 dimensional templates grid equally spaced of $10$ [rad] around the injected signals parameters. The figures also shows the fit of a 4dof $\chi^2$ distribution. }
\label{fig:S4}
\end{figure}

\begin{figure}[h!]
\includegraphics[width=0.45\textwidth]{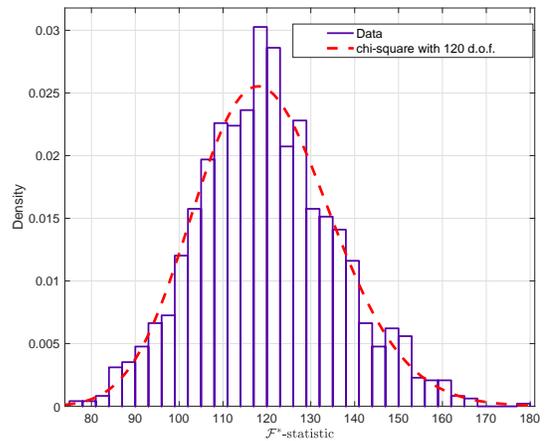}
\caption{Histogram of the detection statistic obtained for a semi-coherent search using a 8 dimensional templates grid equally spaced of $10$ [rad] around the injected signals parameters. The figures also shows the fit of a 120dof $\chi^2$ distribution that  match the experimental histogram.}
\label{fig:S120}
\end{figure}

After the phase metric $g_{ij}^\varphi$ has been inverted into the new phase variables $\Phi$, we need to check that the new metric $g_{ij}^\Phi$ estimates correctly the distance at which the templates  produce a mismatch  $~<1$ rad, we will call this distance  `` mismatch length''.  Using the $\Phi$ variables we have seen that $m_f<1$ when $|\vec{\Delta \Phi}|>1$. So in the $\Phi$ space the mismatch length will be given between two templates separated by more of $|\vec{\Delta \Phi}|=1$ rad. Practically we are asking that the templates are nearly orthogonal to each other (see Appendix \ref{ap:D} for more details).   A  possible way to test this is to run the analysis for a point in the parameter space distant from the injected signal more than the mismatch length. In the case that the mismatch length is estimated correctly and a signal is present in gaussian noise, we expect that the outcome of an analysis performed with a template grid with spacing much larger than the mismatch length, will result as computing the detection statistic for different noise realization. We have created simulated gaussian noise with a software injection signal with SNR $ ~10$ (referred to 1 month of coherent integration) and we have computed and histogrammed the detection statistic for a template grid spaced more than the mismatch length.  The spacing of the grid was about $\Delta \Phi_i=10$ rad for each phase variable. Figure \ref{fig:S4} shows the histogram of the detection statistic obtained. As expected, in the case of a full coherent search, the detection statistic is a 4 dof $\chi^2$ if only gaussian noise is entering into the analysis though the matched filter. We have also performed the same check using a semi-coherent search performed with 30 chunks of data. In this case we expect a $\chi^2$ with 120 dof as Fig. \ref{fig:S120} shows.

\subsection{Fraction of signal-to-noise ratio loss:} 
\begin{figure*}[htp!]
\centering
\includegraphics[width=1.1\textwidth]{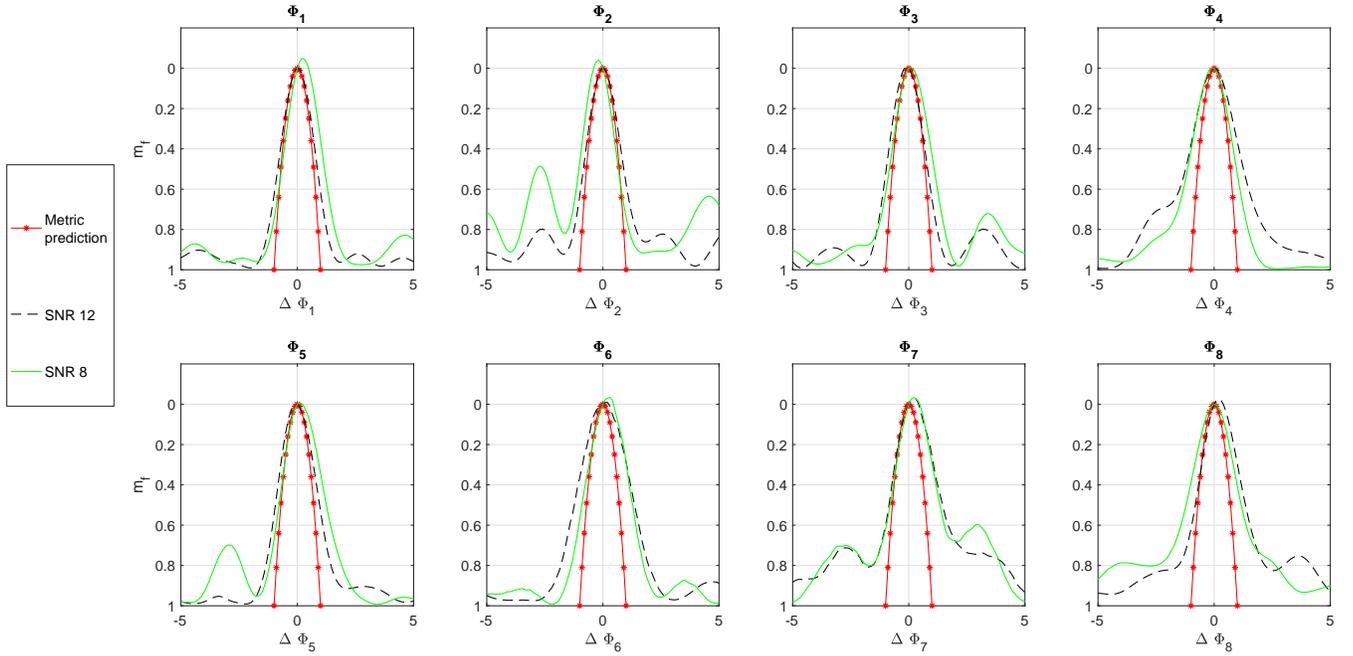}
\caption{Mismatch function $m_f$ computed for templates with a mismatch $\Delta \Phi$. The red dotted line represent the quadratic approximation given by the template metric, while the different line types represent the observed behaviour of the mismatch function for  two software injections with different SNR in gaussian simulated data.}
\label{fig:mism8}
\end{figure*}

The last test is to check in which limit the new variables $\Phi$ and the new phase metric $g^{\Phi}_{ij}$ efficently approximate the mismatch of Eq. \eqref{eq:mism}. Hence we should check if the metric efficently tells us which is the fraction of the signal that we are recovering in our analysis given a template mismatch $\Delta \Phi_i$. Usually Eq. \eqref{eq:mism} is well approximated by the metric for mismatches $<0.5 \%$ \cite{Wette2016} because far away from the signal's true parameter the second order expansion is no more sufficient. Figure \ref{fig:mism8} shows the mismatch function in Eq. \eqref{eq:mism} computed for different software injections with different signal-to-noise ratios with respect to the  mismatched variables $\Delta \Phi_i$. The red dotted curve represents the fraction of SNR loss predicted by the metric, as we can see from the simulation, the signal is completely lost for mismatches$|\Delta \Phi_i| > 1$, as we expect. The secondary modes in each plot are due to noise contributions  or to secondary peaks due to the sidereal responses which are not taken into account in our maximum phase decomposition. Figure \ref{fig:mism8} also points us to another drawback of using more variables than what are needed. In principle, it is possible to have a template in the eight-dimensional parameter space which fit better than the  injected  one in the four-dimensional parameter space. 
However by working directly in the eight-dimensional parameter space (without coming back), this is not a problem, since all the templates that are within a distance  of  $|\Delta \Phi<|1$ from each other count as the same template under the point of view of a mathched filter grid.  

\section{Applications \label{sec:5}}

In the previous sections we have shown how to define a metric in which the mismatch function and hence the likelihood surfaces can be studied . Indeed the topology of the statistic is a feature introduced in the data from the presence of a CW signal  that during the analysis will be matched using certain template grids. It follows that signal that does not have the phase evolution in  Eq. \eqref{eq:phevol} will not show the expected topology of a CW signal. 

We have also seen that using the maximum phase decomposition $\Phi$, the metric $g_{ij}^\Phi$ (and hence geometry of the statistic) can be approximated as an identity matrix. The characteristic geometry in the statistic introduced by $g_{ij}^\Phi$ can be used to try to distinguish between the presence of a CW signal or the presence of non-stationary noise artifacts.  Different types of application can be found, but in this paper we will present 3 different test cases in which the $\Phi$ variables can help for the identification and detection of a CW signal.

\subsection{Frequentist p-values} 
\begin{figure}[h!]
\includegraphics[width=0.5\textwidth]{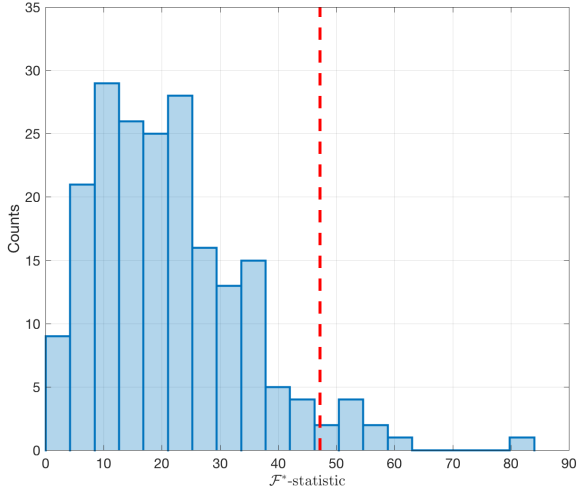}
\caption{Histogram of the noise-only distribution obtained drawing samples with a template spacing $\Delta \Phi=10$ rad. The red vertical dashed line show the value of the detection statistic obtained from an outlier due to a known noise line in O1 data. Its original p-value was about $10^{-6}$ and now is $0.04$}
\label{fig:pf}
\end{figure}

Let us assume that we have obtained some interesting outliers \footnote{With outliers we mean points in the parameter space $\vec{\lambda}$ which show a false alarm probability below a given threshold and need deeper studies or could be due to a real CW signal.} from a given search (semi-coherent or full-coherent). In order to better estimate the significance of the outlier one usually want to use the noise-only distribution of the detection statistic. This distribution is analytically known just in the case only gaussian noise is present together with the signal. More importantly one would like also to capture non-gaussianities inside data and take them into account when computing the p-values. The modelization of non-gaussian noise cannot be done directly, since we do not perfectly known the noise of the experiment. Instead, we  can try to build empirically the noise only distribution by performing the analysis for templates in which we assume that no-signal is present. For example in the case that our only parameters are $f_0$ and $\dot{f}_0$, one can run several analysis spaced more than the frequency and spin-down bins,  thus obtaining different noise-realization (if we assume the ergodic principle) and later build the noise-only distribution with the obtained samples. Two requests must be satisfied when following this procedure: {\it (i)} templates should be far enough from the signal in such a way to blind our analysis to its presence; {\it (ii)} in order to preserve the noise properties the templates should not be too far from a given interesting CW candidate. The $\Phi$ variables give us a clear framework in which the previous constrains are satisfied. In fact, if a template is distant from the signal $\Phi$ parameters by $ |\Delta \Phi|>1$, then we expect to not see anymore the contribution of the signal. Using such kind of technique to generate the noise background can be seen as we are answering the question `` {\it Which is the probability that modulating the noise with a phase modulation very similar to the one of a possible signal but independent, the noise will mimic the GW antenna pattern for which I am looking for?}''. 
In Fig. \ref{fig:pf} we show an example of significance assignement for an outlier due to a known noise line in detectors data found in the last narrow-band search for CW from the pulsar J1952+3252 using O1 data \cite{PhysRevD.96.122006}. The outlier displayed a very high significance (p-value$=10^{-6}$) from the narrow-band search, while generating the noise background with the $\phi$ variables and a template spacing of $\Delta \Phi=10$ we have drawn samples from the noise only distribution obtaining a new sub-threshold p-value for the outlier of $0.04$. 
The fact that  the p-value is increased from $10^{-6}$ to $0.04$ is an indication of the fact that  there are non-gaussian noise that is entering into the analysis and the outlier is likely due to this noise contribution.

\begin{figure*}[htp]
\includegraphics[width=1\textwidth]{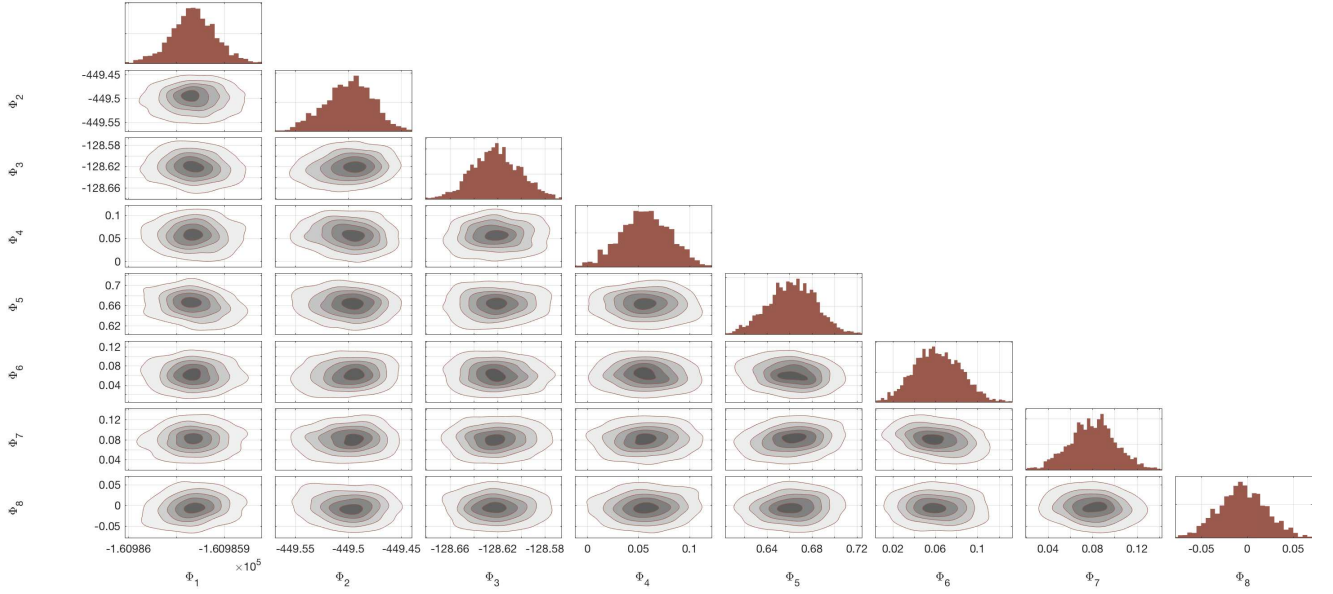}
\caption{Marginalized likelihood of $\mathcal{L}\big(h(\vec{\Phi})|x\big)$ obtained for a full-coherent search  of the hardware injection Pulsar 3 in O1 data (with one month of coherent analysis). The confidence intervals clearly show spherical symmetry with respect to a central point.}
\label{fig:8hs}
\end{figure*}
\subsection{Bayesian Confidence intervals}  
The geometry of the $\overset{_*}{\mathcal{F}}-statistic$ with respect to the intrinsic parameters $\Delta  \vec{\lambda}$ will have an impact on the credible intervals for an analysis based on Bayesian inference. In fact,  by using Eq. \eqref{eq:RML} with Eq. \eqref{eq:mism1} and Eq. \eqref{eq:mism} we have:
\begin{equation}
\mathcal{L}_{\rm ML}(x|h(\Delta \vec{\lambda}))=e^{\overset{_*}{\mathcal{F}}_s[1-g_{ij} \Delta \lambda_i \Delta \lambda_j]}
\end{equation}
being $\overset{_*}{\mathcal{F}}_i$ the statistic associated to the matched parameters of the signal. If we use the phase variables it is easy to see that:
\begin{equation}
\mathcal{L}_{\rm ML}(x|h(\Delta \vec{\lambda}))=e^{\overset{_*}{\mathcal{F}}_s}e^{-\overset{_*}{\mathcal{F}}_s \sum_{k=1}^8 \Delta \Phi_k^2} 
\label{eq:30}
\end{equation}

\begin{figure*}[htp]
\includegraphics[width=1\textwidth]{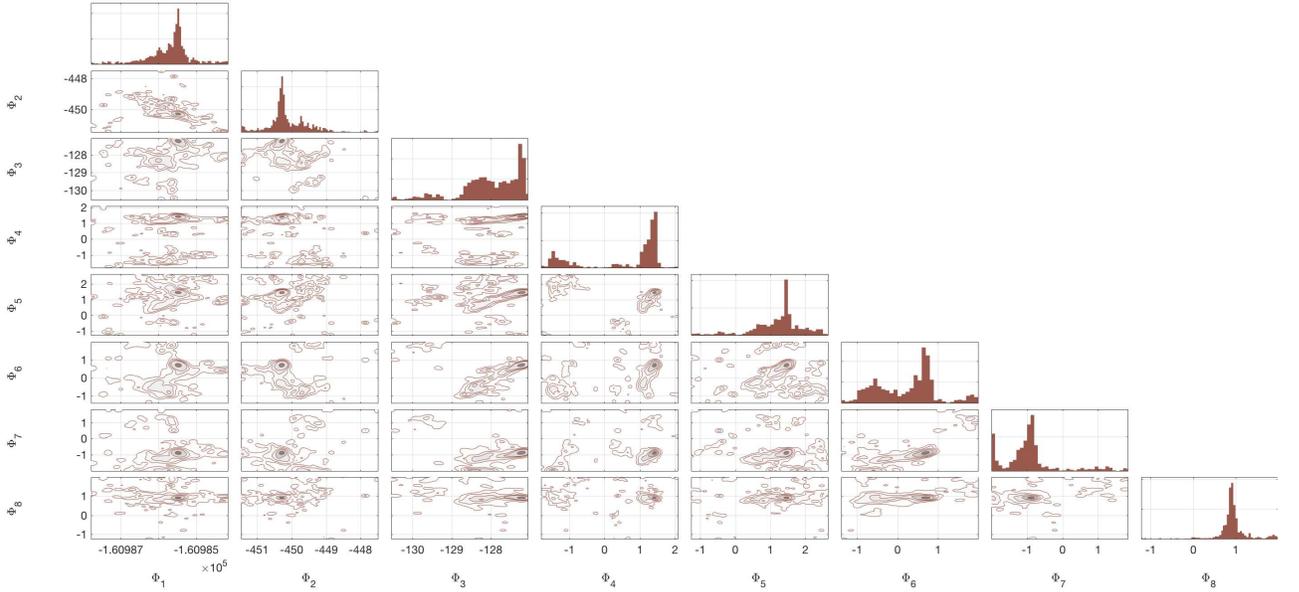}
\caption{Marginalized likelihood of $\mathcal{L}\big(h(\vec{\Phi})|x\big)$ obtained for a full-coherent search performed in the case of software simulated gaussian data with a monochromatic noise line injected. The posteriors distribution are clearly not characteristic of a CW signal.}
\label{fig:8hn}
\end{figure*}

 For very strong signals, we expect the maximum likelihood estimator to be a $\delta$-like function around the signals parameters, while for low signal-to-noise ratio we expect the posteriors to be more similar to gaussians always centered around the signal parameters. We can also study the confidence intervals with respect to the $\Phi$ variables:
\begin{equation}
\int_{\Omega(\Phi_s)} \mathcal{L}_{\rm ML}(x|h(\Delta \vec{\lambda}) d \vec{\Phi}=0.95,
\end{equation}
where $\Omega(\Phi_s)$ is a given volume in the parameter space centered around a value $\Phi_s$ that can be the mean of the maximum likelihood estimator. From Eq. \eqref{eq:30} it is clear that the maximum likelihood estimator has spherical symmetry with respect to the templates computed in the $\Phi$ space. For example, Fig. \ref{fig:8hs} shows the contour plots obtained running a Markov Chain Monte Carlo algorithm looking for a software injected signal with signal-to-noise ratio 8 in one month of O1 data. It is clear from the figure that the posterior have spherical symmetry as we expect.  On the other hand Fig. \ref{fig:8hn} shows the contour plots obtained by running the same algorithm for a very loud (signal-to-noise ratio about 300) monochromatic noise line injected at the frequency searched in the analysis, in software simulated gaussian data. It is clear that in this case the posterior distribution has not spherical symmetry. We can qualitatively use this for distinguish among CW signals and non-gaussian noise lines. For example one can compute the marginalized probability $p(r,r_c)$ to be in a spherical volume $\mathcal{S}(r,r_c)$ from a central point $r_c$.
\begin{equation}
p(r,r_c)=\int_{\mathcal{S}(r,r_c)}  \mathcal{L}_{\rm ML}\big(x|h(\Phi \big) d\vec{\Phi}
\end{equation}
In the case of  a CW signal, we expect $p(r,r_c)=1$ if the radius of the sphere is within one template space ($\Delta \Phi<1$). For noise-lines, instead, since the spherical symmetry is not preserved and the posterior is spread all over the template grid, we expect $p(r,r_c)$ to not increase so rapidly from the central point $r_c$ and to reach the value of $1$ for $\Delta \Phi>1$. Table \ref{tab:1} reports this kind of test performed for the examples in Fig. \ref{fig:8hs} and Fig. \ref{fig:8hn}.
\begin{table}
\caption{l\label{tab:1}Second column: radius of the spherical volume in the parameter space $\Phi$ for which we compute $p(r,r_c)$ . Third column: Value of $p(r,r_c)$. }
\begin{tabular}{c|c|c}
\textbf{Case} & radius [deg] & $p(r,r_c)$ \\
\\
\hline
Signal  & 0.05 & 0.2272\\
Signal  & 0.1 & 0.97\\
Signal & 0.15 & 1.0\\
Noise line & 1.0 & 0.0042\\
Noise line & 1.5 & 0.2676\\
Noise line & 2.5 & 0.7366 \\
\end{tabular}
\end{table}
\

\begin{figure}[h!]
\includegraphics[width=0.45\textwidth]{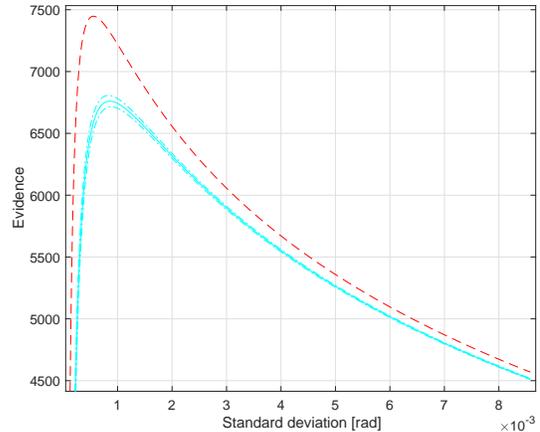}
\caption{Evidence (vertical axis) computed with respect to a chosen variance (horizontal axis), in the hypothesis of a multivariate gaussian distribution for the $\Phi$ variables. Red dashed line: Evidence trend for O1 data around the Hardware Injection Pulsar 3. Blue solid line: Evidence trend for gaussian samples generated with a variance equal to the inverse of the maximum statistic found in the search, the lines cover the $1 \sigma$ confidence interval (blue dotted lines). }
\label{fig:p8z}
\end{figure}

\begin{figure}[h!]
\includegraphics[width=0.45\textwidth]{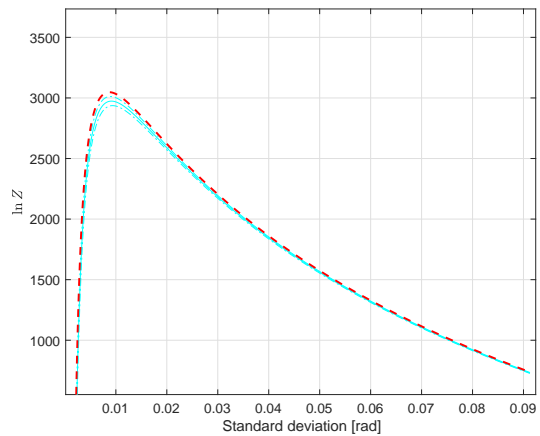}
\caption{Evidence (vertical axis) computed with respect to a chosen variance (horizontal axis), in the hypothesis of a multivariate gaussian distribution for the $\Phi$ variables.  Red dashed line: Evidence trend for a software injected signal with SNR 8. Blue solid line: Evidence trend for gaussian samples generated with a variance equal to the inverse of the maximum statistic found in the search, the lines cover the $1 \sigma$ confidence interval (blue dotted lines).}
\label{fig:8z}
\end{figure}

\begin{figure}[h!]
\includegraphics[width=0.45\textwidth]{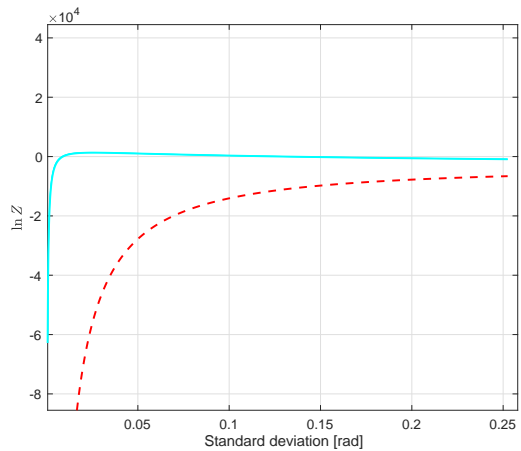}
\caption{Evidence (vertical axis) computed with respect to a chosen variance (horizontal axis), in the hypothesis of a multivariate gaussian distribution for the $\Phi$ variables.  Red dashed line: Evidence trend for a monochromatic noise line injected in gaussian generated noise. Blue solid line: Evidence trend for gaussian samples generated with a variance equal to the inverse of the maximum statistic found in the search, the lines cover the $1 \sigma$ confidence interval (blue dotted lines). }
\label{fig:nz}
\end{figure}

A more quantitative way to check the spherical symmetry hypothesis is by using the evidence $Z$ (or marginalized likelihood) of having $N_s$  samples from a multivariate normal 8-d distribution. 
In fact, according to Eq. \eqref{eq:30} if we run a Markov chain monte carlo (MCMC) for the  maximum likelihood estimators  $\mathcal{L}_{\rm ML}(x|h(\Delta \vec{\lambda})$ what we expect to see are roughly samples from a eight-dimensional bivariate normal distribution with mean the parameter of the signal and variance the $\sigma^2 \approx 1/ \overset{_*}{\mathcal{F}}_s$.
For a  8 dimensional multivariate normal distribution with mean $\vec{\mu}$ and variance $\sigma^2$, the logarithm of the evidence can be computed as:
\begin{equation}
\ln Z= -4 N_s \ln(2 \pi)-\frac{1}{2} N_s \ln (\sigma^2) - \frac{1}{2} \sum_{i=1}^{N_s} \frac{(\vec{x}_i-\vec{\mu}) \cdot (\vec{x}_i-\vec{\mu})}{\sigma^2}.
\label{eq:evidence}
\end{equation}
The idea is to run a  Markov chain monte carlo (MCMC) algorithm for the maximum likelihood estimator and then from the output evaluate the mean $\vec{\mu}$, then compute the evidence for many values of $\sigma^2$.
We expect the evidence to have a peak in correspondence of the value $\sigma^2_s=1/\overset{_*}{\mathcal{F}}_s$ (according to \eqref{eq:30} and then we expect a linear decrease.  
Practically this means that the samples obtained from the MCMC should be representative of an 8-dimensional gaussian process. 
We then perform the following procedure to  probe the nature of data: {\it (i)} We run a MCMC algorithm on the data in order to sample the maximum likelihood estimator in Eq. \eqref{eq:30}. {\it (ii)} After obtaining $N_s$ independent samples we compute the mean $\vec{\mu}$ and the variance $\sigma^2_s=1/\overset{_*}{\mathcal{F}}_{\rm max}$ of the distribution, where we used the maximum of the $\overset{_*}{\mathcal{F}}$-statistic found by the MCMC. {\it (iii)} Using several values of $\sigma^2$ we compute the logarithm of the evidence $\ln Z_{\rm data}$ in Eq. \eqref{eq:evidence} as function of the variance, we expect to see a peak around $\sigma^2_s$ and after that a linear decrease. {\it (iv)} As another proxy for the evidence $\ln Z_{\rm data}$ to be representative of a gaussian process, we software generate $N_s$ samples of  a 8-dimensional gaussian process with given mean $\vec{\mu}$  and variance $\sigma^2_s$, we then compute the evidence $\ln Z_{\rm proxy}$ as a function of the variance.   {\it (v)} The two evidences  $\ln Z_{\rm data}$ and $\ln Z_{\rm proxy}$ are compared together.
 If the evidence curve for the data is above or within the a degree of uncertainty (given by the statistical standard deviation of $Z$ for a pure gaussian process) of  the evidence curve generated by true gaussian samples,  then we have a strong reason to believe that what we are observing is likely due to a signal in Gaussian noise.  
 Figs. \ref{fig:p8z}, \ref{fig:8z}, \ref{fig:nz} show the evidence computed from a Markov chain Monte Carlo ran to sample the maximum likelihood estimators of the hardware injection Pulsar 3 in one month of O1 data (SNR~70), a software injection with SNR 8 (with the same parameters of Pulsar 3 but in gaussian simulated data) and a monochromatic noise line injected in gaussian simulated data with an high SNR that contaminates the analysis. 
In all the figures, the evidence computed from the data is compared between the proxy evidence computed from software generated samples of a bivariate normal distribution.
The figures show that in the case that a signal is present inside data, the evidence curve of the data is above or within the evidence generated by software generated from gaussian process with same variance and mean. In the case of Pulsar 3 (high SNR) we observe that the evidence of the data is above the evidence generated by the software generated gaussian process, meaning that the recovered likelihood is more ``peaked''  than the one expected. This is reasonable since the signal is very strong and we are neglecting the effect of sidereal modulations which can further modify the shape of the likelihood surfaces in many different local peaks. Figure \ref{fig:nz} instead shows the evidence computed in the case a very strong monochromatic noise line is present inside the data. It is clear that the evidence of data is far below the evidence computed for a gaussian process with same variance and mean, meaning that the likelihood that we are observing has not spherical symmetry at all and hence is very unlikely to be generated by a signal in gaussian noise. 

\subsection{Application to Markov Chain Monte Carlo Follow-up:} Another possible application of the new variables is in the so-called follow-up algorithms. Follow-ups are procedures aimed to understand the nature of a given candidate. Depending on the necessities of the problem usually we want these algorithms to follow candidate in the parameter space in such a way to perform longer and longer searches in order to increase the significance of a possible CW detection. Recently the possibility of performing this tasks with Markov Chain Monte Carlo Techniques have been shown \cite{Ashton2018}.  It is well known that Markov Chain Monte Carlo should be tailored on the type of posterior that we would like to sample. Using the $\Phi$ variable the geometry of the posterior $p\big(h(\Phi)|x\big)$ is well known and this may help the algorithm to converge faster thus saving computation time. Running the same algorithm used in \cite{Ashton2018} from band sub-sampled data \cite{PiccinniPhD},  but implemented for the $\Phi$ variables, we have found an Integrated Autocorrelation time\footnote{The Integrated Autocorrelation is an estimator of how many samples are necessary in order to have two independent samples from a MCMC algorithm \cite{Sokal1996}.} of about $25$ iterations whereas in the original work it is about $90$ iterations. Thus meaning that  about the half of the iterations are needed in order to obtain the posterior distribution, even if we are using 4 additional variables. The decreased computational cost grant us the possibility to increase the sensitivity of the search by increasing the number of outliers to follow-up.

\section{Conclusion \label{sec:6}}

In this paper we have presented a new set of variables for CW, called maximum phase decomposition, which is able to regularize the template metric for semi-coherent and full coherent searches. As shown in the paper the template metric plays a very important role when deciding how to build the template lattice. The fact that, in the standard CW searches, the metric for long coherent searchers is handle numerically makes the template grid non-trivial to built. At the current state the maximum phase decomposition cannot be used to to built a template grid for all-sky or semi-coherent searches due to the fact that it will bring to the placement of many templates in the eight-dimensional template space which does not have a correspondent template in the four-dimensional parameter space.

However, applications for new variables $\Phi$ may result in more practical tasks for CW searches such as studies on the significance and nature of candidates using either frequentist or Bayesian frameworks. In particular, the $\Phi$ variables makes possible to study and explore the response of the noise to templates which have a very similar phase modulation (even if slightly non-physical) to a given CW candidate.
A focal application of the new $\Phi$ variables can take place in Markov Chain Monte Carlo follow-ups, in fact the usage of such new variables can significantly improve the convergence of the algorithm thus reducing computation time. This means that we will able to follow-up more candidates thus improving the sensitivity of our searches. Even though physical parameters cannot be recovered at the end of the follow-up, the maximum phase decomposition still offers a good tool to study the significance of the CW candidate when increasing the coherent integration.  The implementation of this framework for this kind of algorithm will be presented in a future work.
Summarizing the maximum phase decomposition offer a valuable and alternative tool to exploit the phase properties of CW signal and to study the response of the detector noise to such phase modulations.

\begin{acknowledgments}
The authors would like to thank Gregory Ashton, Karl Wette and Matthew Pitkin for the useful comments during the development of this work. Simone Mastrogiovanni would like to thank also the University of Rome Sapienza that has supported this work with the action ``Avvio alla Ricerca 2017''.
This paper carries LIGO Document Number \dcc. 
\end{acknowledgments}

\onecolumngrid
\appendix
\section{The Phase metric computation \label{ap:A}}
Let us assume that the data is a superposition of gaussian noise and a CW signal: 
\begin{equation}
\ket{h}=H_0 \sum_s H_s \ket{\mathcal{A}_s},
\end{equation}
were we have dropped the dependence from $\vec{\lambda_s}$ for the notation sake, and that we are performing our analysis using a set of templates computed for a small mismatch $\Delta \vec{\lambda}$. If we assume the data to be a superposition of gaussian noise and signal, like in Eq. \eqref{eq:super}, and taking the definition of $\overset{_*}{\mathcal{F}}$-statistic in Eq. \eqref{eq:R} we can write: 
\begin{equation}
\overset{_*}{\mathcal{F}}_{mismatch}=\frac{1}{2} \sum_p \frac{(n^*_p+ \sum_s H^*_s \braket{\mathcal{A}_s|\mathcal{A}_p} )( n_p+ \sum_s H_s \braket{\mathcal{A}_p|\mathcal{A}_s} )}{\braket{\mathcal{A}_p|\mathcal{A}_p}}. 
\end{equation}
Being $n_p$ the projection of the noise over the template and  `` $*$'' the complex conjugator operator. We can now take the expected value of the above equation and exploit the products obtaining:
\begin{equation}
 E[\overset{_*}{\mathcal{F}}_{mismatch}]=\frac{1}{2} \sum_p \frac{E[n^*_p \cdot n_p]}{\braket{\mathcal{A}_p|\mathcal{A}_p}}+ \frac{E[n^*_p \sum_s H_s \braket{\mathcal{A}_p|\mathcal{A}_s}]}{\braket{\mathcal{A}_p|\mathcal{A}_p}}+ \frac{E[n_p \sum_s H^*_s \braket{\mathcal{A}_s|\mathcal{A}_p}]}{\braket{\mathcal{A}_p|\mathcal{A}_p}}+ \frac{E[ \sum_s H^*_s \braket{\mathcal{A}_s|\mathcal{A}_p} \sum_s H_s \braket{\mathcal{A}_p|\mathcal{A}_s}]}{\braket{\mathcal{A}_p|\mathcal{A}_p}} 
 \label{eq:long}
\end{equation}
The first term in Eq. \eqref{eq:long} represents the contribution from the noise\footnote{The distribution of the statistic in the noise case in a $\chi^2$ with $4 N$ degree of freedom, where $N$ are the number of chunks used in the analysis.}, the second and third terms vanish since we are assuming gaussian noise with zero mean. Finally the last term represent the contribution from a possible overlap of the signal and the mismatched template. Taking into account the previous considerations, we can rewrite Eq. \eqref{eq:long} exploiting the summation over the $s$ index in the last term:
\begin{equation}
E[\overset{_*}{\mathcal{F}}_{m}]= \sum_p  \frac{E[n^*_p \cdot n_p]}{\braket{\mathcal{A}_p|\mathcal{A}_p}}+ \frac{1}{\braket{\mathcal{A}_p|\mathcal{A}_p}}  E \bigg[ \sum_{s_1=s_2} |H_s|^2 |\braket{\mathcal{A}_s|\mathcal{A}_p}|^2 + \sum_{s_1 \neq s_2} H_{s_1} H^*_{s_2} \braket{A_{s_2}|\mathcal{A}_p} \braket{\mathcal{A}_p|\mathcal{A}_{s_1}}\bigg] \\
\label{eq:caos}
\end{equation}
Let us focus on the last term in Eq. \eqref{eq:caos} which represent the contribution to the detection statistic from possible interference of two different polarization of a CW signal.

In the limit of long integration time (greater than 1 day), the antenna response function to the two polarization $+$ and $\times$ are expected to become almost independent each other \cite{Jaranowski1998}, this happens because the antenna response is averaged over a long integration time. We can then write:
\begin{equation}
\frac{E [ \sum_{s_1 \neq s_2} H_{s_1} H^*_{s_2} \braket{\mathcal{A}_{s_2}|\mathcal{A}_p} \braket{\mathcal{A}_p|\mathcal{A}_{s_1}}]}{\braket{\mathcal{A}_p|\mathcal{A}_p}}=E [ \sum_{s_1 \neq s_2} H_{s_1} H^*_{s_2} \braket{\mathcal{A}_{s_2}|\mathcal{A}_{s_1}}]\approx 0
\label{eq:est}
\end{equation}
Equation \eqref{eq:est} is telling us that the contribution from the interference of two different polarization vanish in the limit that the signal is formed by orthogonal polarization. Finally Eq. \eqref{eq:caos} can be rewritten in the more compact form:
\begin{equation}
E[\overset{_*}{\mathcal{F}}_{m}]=\sum_p  \frac{E[n^*_p \cdot n_p]}{\braket{\mathcal{A}_p|\mathcal{A}_p}}+ \sum_{s} |H_s|^2 |\mathcal{A}_s|^2 \frac{ |\braket{\mathcal{A}_s|\mathcal{A}_p}|^2}{\braket{\mathcal{A}_p|\mathcal{A}_p} \braket{\mathcal{A}_s|\mathcal{A}_s}} = \frac{E[n^*_p \cdot n_p]}{\braket{\mathcal{A}_p|\mathcal{A}_p}}+ \sum_{s} \overset{_*}{\mathcal{F}}(\beta_s,\lambda_s) M_{sp}(\lambda_s,\lambda) \\
\label{eq:R_m}
\end{equation}
Where $\overset{_*}{\mathcal{F}}(\beta_s,\lambda_s)$ is the detection statistic for a perfect matched template while the loss due to the template mismatch is encoded in a mismatch matrix that depend on the coupling between the true signal polarization $s$ and the template polarizations $p$.
\begin{equation}
M_{sp}(\lambda_s,\lambda)=\frac{ |\braket{\mathcal{A}_s|\mathcal{A}_p}|^2}{\braket{\mathcal{A}_p|\mathcal{A}_p} \braket{\mathcal{A}_s|\mathcal{A}_s}}.
\label{eq:brak}
\end{equation}
The mismatch matrix can be easily computed in the time basis, remembering that in general a template can be written in the time basis as in Eq. \eqref{eq:tb}:
\begin{equation}
M_{sp}=\frac{1}{T_{\rm coh}^{2}} \bigg| \int_0^{T_{\rm coh}} e^{i \Delta \phi_{sp}(t;\lambda_s,\lambda)} dt \bigg|^2
\label{eq:msp}
\end{equation}
Being $ \Delta \Phi_{\rm sp}(t;\lambda_s,\lambda)$ the phase mismatch between two polarizations $s$ and $p$ of the signal and template.  Finally let us assume that the signal we are looking for is composed by the usual $+$ and $\times$ polarization. If we take the mismatch function from Eq. \eqref{eq:mism} and for $ E[\overset{_*}{\mathcal{F}}_{m}]$ we use  Eq. \eqref{eq:R_m}:
\begin{equation}
m_f(\lambda, \lambda_s)=1- \frac{\overset{_*}{\mathcal{F}}_+ (\beta_s,\lambda_s) [M_{++} (\lambda,\lambda_s)+M_{+ \times} (\lambda,\lambda_s)+\overset{_*}{\mathcal{F}}_\times (\beta_s,\lambda_s) [M_{\times \times} (\lambda,\lambda_s)+M_{\times +} (\lambda,\lambda_s)]}{\overset{_*}{\mathcal{F}}_\times (\beta_s,\lambda_s)+\overset{_*}{\mathcal{F}}_+ (\beta_s,\lambda_s)}.
\label{eq:aaa}
\end{equation}
We can now Taylor expand up to the second order the matrix $M_{\rm sp}$ around the parameter of the signal (a summation over the polarization indexed $s$ and $p$ is intended). If we assume  $\overset{_*}{\mathcal{F}}_+ (\beta_s,\lambda_s) \approx \overset{_*}{\mathcal{F}}_\times (\beta_s,\lambda_s)$ \footnote{Which is a reasonable assumption since the sidereal responses are averaged over a very long integration time.}
\begin{equation}
m_f(\lambda,\lambda_s)= 1- \frac{1}{2} \bigg[M_{sp}\bigg|_{\lambda=\lambda_s}+\sum_j \frac{\partial M_{sp}}{ \partial \lambda_j} \bigg|_{\lambda=\lambda_s} \Delta \lambda_j +\sum_{j,1} \frac{1}{2}\frac{\partial M_{sp}}{ \partial \lambda_j \partial \lambda_i} \bigg|_{\lambda=\lambda_s} \Delta \lambda_i \Delta \lambda_j+\mathcal{O}(\Delta \lambda^3)\bigg] = g_{ij} (\lambda_s) \Delta \lambda_i \Delta \lambda_j \mathcal{O}(\Delta \lambda^3)
\label{eq:long}
\end{equation}

The terms given by the first derivatives in Eq. \eqref{eq:long} are zero, since we are expanding around a local maximum of the mismatch function. From the definition in Eq. \eqref{eq:brak} it is possible to see that $0-th$ terms are $M_{++}(\lambda_s)=M_{\times \times}(\lambda_s)=1$ and the terms $M_{+\times}(\lambda_s)=M_{\times +}(\lambda_s)=0$. If one computes the remaining second order derivatives,
In the case that the phase mismatch due to the sidereal templates is negligible with respect to the phase mismatch introduced by all the other modulations  and that we are integrating over many cycles of the signal\footnote{This is often a reasonable assumption for long integration time, since effect such as the Romer delay are always bigger and for signals in the range from the Hz to the kHz.}, Equation \eqref{eq:long} will be finally reduced to the form:
\begin{equation}
m_f(\lambda,\lambda_s)=  \sum_{i,j}\bigg[\frac{1}{T_{\rm coh}} \int_0^{T_{\rm coh}}  \frac{\partial \phi}{\partial \lambda_i} \frac{\partial \phi}{\partial \lambda_i}  \bigg|_{\lambda=\lambda_s} dt- \frac{1}{T^2_{\rm coh}}  \int_0^{T_{\rm coh}}   \frac{\partial \phi}{\partial \lambda_i}  \bigg|_{\lambda=\lambda_s} dt  \int_0^{T_{\rm coh}}   \frac{\partial \phi}{\partial \lambda_j}  \bigg|_{\lambda=\lambda_s} dt \bigg]\Delta \lambda_i \Delta \lambda_j+\mathcal{O}(\Delta \lambda^3).
\label{eq:metric_final}
\end{equation}
Where the polarization indexes $s,p$ are no more present since we are neglecting the phase modulations of the sidereal motion with respect to other modulations. In Eq. \eqref{eq:metric_final} one can recognize the form of the phase metric presented in Eq. \eqref{eq:phase_metric}.
\twocolumngrid

\section{Semi-coherent metric \label{ap:B}}
In semi-coherent searches such as \cite{2015ApJ...813...39A} the data is split in several data chunks of nearly same duration. The matched filter is then applied to each data chunks obtaining a value of the statistic $\overset{_*}{\mathcal{F}}$ for each data chunk $l$ and later combined incoherently. In practice the final value of the statistic will be the summation of all the obtained values. For our search, we define the mismatch as:
\begin{equation}
m_f=\frac{\sum_l^{\rm chunks} \overset{_*}{\mathcal{F}^l_s} -  \overset{_*}{\mathcal{F}^l_m}}{\sum_l^{\rm chunks} \overset{_*}{\mathcal{F}^l_s}},
\end{equation}
where $s,m$ refers to the expected values of the statistic for the signal parameters and for a set of mismatched parameters. Following the same procedure of Appendix \ref{ap:A} we can Taylor expand the up to the second term in order to obtain the metric:
\begin{equation}
m_f=\frac{\sum_l^{\rm chunks} \overset{_*}{\mathcal{F}}^l_s g_{ij}^{l} \Delta \lambda_i \Delta \lambda_j }{\sum_l^{\rm chunks} \overset{_*}{\mathcal{F}}^l_s}
\end{equation}
 where $g_{ij}^{l} $ is the phase metric in Eq. \eqref{eq:phase_metric} computed for the chunk $l$. If the data is split into chunks of the same length we expect (in a case of a CW) $\overset{_*}{\mathcal{F}}^l_s$ to be almost the same over each data chunk $l$. We can then simplify the above equation as
\begin{equation}
m_f \approx \frac{\sum_l^{\rm chunks} g_{ij}^{l} \Delta \lambda_i \Delta \lambda_j }{N},
\end{equation}
where $N$ is the number of data chunks. It follows that the semi.-coherent metric  $\tilde{g}_{ij}$ can be defined as
\begin{equation}
\tilde{g}_{ij}= \frac{\sum_l^{\rm chunks} g_{ij}^{l}}{N}.
\end{equation}
This expression is equivalent to the one already found in \cite{Wette2015} in which the author have proven all the approximations done.

\section{Effect of extra-dimensions \label{ap:C}}

\begin{figure}[h!]
\includegraphics[width=0.4\textwidth]{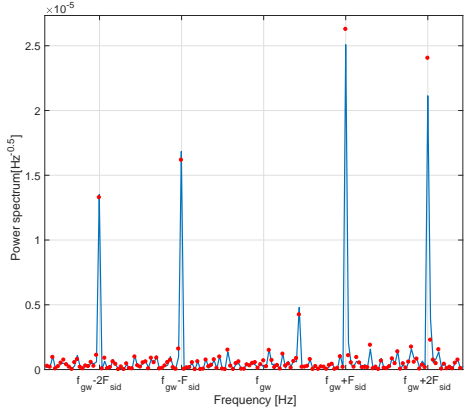}
\caption{Red-line: Power spectrum of the O1 data (1 month of integration) corrected for the $\Phi$ parameters associated with the Hardware injection Pulsar 3. The 5 frequency peaks due to the sidereal modulation are clearly visible and the associated $\mathcal{F}$-statistic was ~1150. Blue line: Power spectrum of recovered from the $\Phi$ parameters found in a MCMC search, the reached $\mathcal{F}$-statistic was 1180.}
\label{fig:spec}
\end{figure}

The fact that we are using  $8$ dimensions instead of $4$ is introducing in the analysis  means that in the template lattice there may exists points which have no correspondent template in the $4$ parameter space but may fit better than the original astrophysical template. For instance one can obtain a combination of $\Phi_{1-8}$ which correspond to a combination of the physical phases $\phi_{1-8}$ where  the four parameters $f_0, \dot{f}_0, \alpha, \delta$ does not have the same value, but a slight different value for all the $\phi_{1-8}$ .  Under a point of view of the signal frequency components, this correspond to have more possible combination of the amplitudes on the 5 frequency components of the signal. As an example, Figure \ref{fig:spec} shows two different power spectrum obtained looking for the hardware  injection Pulsar 3  in one month of O1 data. The injected template (which is represented by the blue solid line) clearly shows the 5 frequency peaks due to the sidereal modulation of the signal. The red dashed line on the other hand shows the spectrum obtained for the template built from the $\Phi$ that has a statistic bigger than the original one. The template which maximize the statistic in the 8 dimensional space does not have a correspondent template in the 4 dimensional space.  As we can see, the usage of the $8$ variables $\Phi$ is leaving more degrees of freedom to adjust the polarization components. 

\begin{table}
\caption{l\label{tab:2} Second column: Value of the detection statistic obtained. Third column: relative error on the amplitude estimation computed as the percentage of amplitude loss. Fourth column: relative error on the $\phi$ parameter computed as $\Delta \Psi/90 {\rm deg}$. Last column: relative error on the $\eta$ parameter computed as $\Delta \eta/2$. }
\begin{ruledtabular}
\begin{tabular}{lcccc}
\textbf{Data-set} & $\overset{_*}{\mathcal{F}}$-Statistic &  $\epsilon_{h_0}$ & $\epsilon_{\psi}$&$\epsilon_{\eta}$  \\
\hline
Physical & 1150 & 6 \% & 0.40\% & 0.9\%\\
non-physical & 1180 & 4\% & 0.02\% & 2.0\%\\
\end{tabular}
\end{ruledtabular}
\end{table}

\section{Correlations of the phase templates \label{ap:D}}

\begin{figure}[h!]
\includegraphics[width=0.5\textwidth]{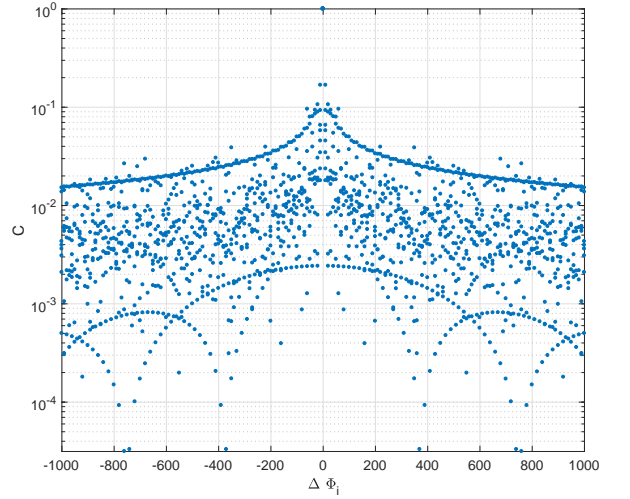}
\caption{Vertical axis: Correlation of two phase templates (one is fixed) with a given distance in the $\Phi$ space (x-axis).}
\label{fig:phase_correlation}
\end{figure}

\begin{figure}[h!]
\includegraphics[width=0.5\textwidth]{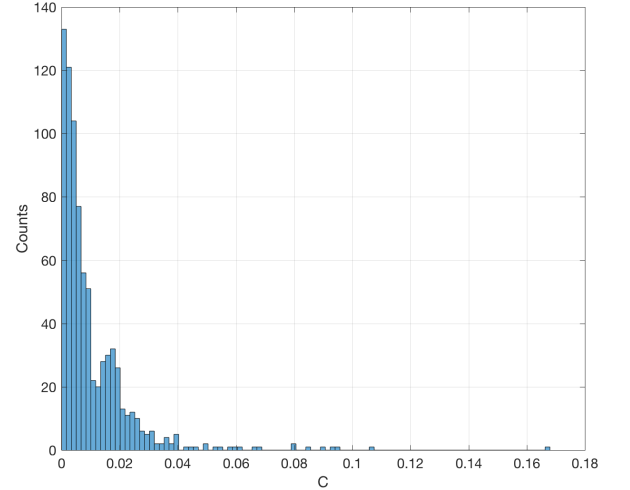}
\caption{Histogram of the correlation values obtained in Fig. \ref{fig:phase_correlation}.}
\label{fig:phase_correlation_histogram}
\end{figure}

As pointed out, we define the correlation among phase templates  as their normalized scalar product:
\begin{equation}
C=\frac{\braket{A_a | A_b}}{|A_a| |A_b|}, 
\end{equation}
where the subscripts $a,b$ indicate two templates computed from the variables $\vec{\Phi}_a$ and  $\vec{\Phi}_b$. Following Eq. \eqref{eq:mism} if the metric estimates correctly the fraction of signal that we are recovering with a template, it follows that two phase templates computed from two parameters s $\vec{\Phi}_a$ and  $\vec{\Phi}_b$ distant $\Delta \Phi=|\vec{\Phi}_a-\vec{\Phi}_b|>1$ will give a correlation very close to zero. Fig. \ref{fig:phase_correlation} shows the correlation of two phase templates computed with a distance in the parameter space that is a multiple of 10. Even though the correlation does not drop immediatly to zero (since we are neglecting the sidereal modulations and since the metric its a quadratic approximation), its value becomes small very fast. One can also plot the histogram of the correlation obtained in this way, that is shown in Fig. \ref{fig:phase_correlation_histogram}, where we see that the majority part of phase templates have a very small correlation with the central point.

\newpage
\bibliography{cw8}

\end{document}